\documentstyle[aps,twocolumn,amssymb]{revtex}
\begin{document}
\newcommand{\be}{\begin{equation}}
\newcommand{\ee}{\end{equation}}
\newcommand{\bq}{\begin{eqnarray}}
\newcommand{\eq}{\end{eqnarray}}
\title{Spin picture of the one-dimensional Hubbard model:
Two-fluid structure and phase dynamics}
\author{Arianna Montorsi and Vittorio Penna} 
\address{Dipartimento di Fisica and Unit\`a INFM,
Politecnico di Torino, I-10129 Torino, Italy}
\date{\today}
\maketitle
\begin{abstract}
We propose a scheme for investigating the quantum dynamics of
interacting electron models by means of a time-dependent variational
principle and spin coherent states of space lattice operators.
We apply such a scheme to the one-dimensional Hubbard model,
and solve the resulting equations in different regimes.
In particular we find that
at low densities the dynamics is mapped into two coupled nonlinear
Schr\"odinger equations, whereas near half--filling the model is
described by two coupled Josephson--junction arrays. Focusing then
to the case in which only the phases of the spin variables are 
dynamically active, we examine a number of different
solutions corresponding to the excitations of few macroscopic modes.
Based on fixed--point equations of the simpler among them, we show
that the standard one--band ground--state phase space is found. 
\end{abstract}

\def\elle{{\it l}}
\def\dsp{\displaystyle}
\def\nks{n_{{\bf k}, \sigma}}
\def\cksd{c_{{\bf k}, \sigma}^{\dagger}}
\def\cks{c_{{\bf k}, \sigma}}
\def\cjs{c_{{\bf j}, \sigma}}
\def\cjsd{c_{{\bf j},\sigma}^\dagger}
\def\nup{n_{{\bf j}, \uparrow}}
\def\ndw{n_{{\bf j}, \downarrow}}
\def\sj{\sum_{\bf j}}
\def\sjk{\sum_{<{\bf j},{\bf k}>}}
\def\sk{\sum_{\bf k}}
\def\skl{\sum_{<{\bf k},{\bf l}>}}
\def\diracjk{\delta_{{\bf j}, {\bf k}}}
\def\sigsig{\delta_{\sigma , \, \sigma '}}
\def\HH{{\cal H}}
\def\numer{{\cal N}}
\pacs{71.10.Fd, 03.65.s, 03.65.B, 47.32.Cc}
\narrowtext
\section{Introduction}
\noindent
Investigating quantum dynamics of strongly correlated many-body
systems is a hard task since, even for extremely simplified models,
the interactions of the large number of degrees of freedom are usually
affected by a nonlinear character.
At the operational level this entails the impossibility of evaluating
explicitly the action of the propagator known from the Schr\"odinger
equation, that is the evolution
$|\Phi \rangle = \exp [-it H/ \hbar]|\Phi_0 \rangle$
of a state $|\Phi_0 \rangle$ governed by the Hamiltonian $H$.
A standard way to reduce such a difficulty to a more tractable form
consists in recasting the purely quantum problem within an appropriate
coherent--states picture once the algebraic structure characterizing
$H$ has been identified. This leads to represent the system evolution
through the equations of motion issued from an effective classical
Hamiltonian $\cal H$ expressed in terms of the coherent state
parameters \cite{ZFG}.

A systematic development of such an approach is provided by
the time-dependent variational principle (TDVP) procedure
\cite{KEKO}. This amounts to constructing a trial macroscopic
wave function $|\Psi \rangle$
that contains time-dependent parameters whose evolution is
derived so as to optimize the approximation of the quantum
propagator action \cite{ELBA}. On this basis, using the generalized
coherent--states to construct the trial state $|\Psi \rangle$ is quite
advantageous in that the coherent state parameters naturally
label $|\Psi \rangle$ and make explicit its dependence on the
algebraic structure of $H$, namely on the operators describing the
microscopic physical processes therein. By making the phase that
appears in $|\Psi \rangle$ coincide with the effective action,
the Schr\"odinger equation turns out to be automatically satisfied
when projected onto $|\Psi \rangle$.

In a recent paper \cite{MOPE} such a scheme was specialized to the
case of interacting electrons described by the Hubbard Hamiltonian.
There the coherent states entering $|\Psi \rangle$ were specific to
the physical regimes ({\sl e.g.}, superconducting, antiferromagnetic,
etc.), the latter selecting case by case the appropriate approximate
algebraic framework within the
Hamiltonian dynamical algebra.

The standpoint here adopted is instead to implement a unified
TDVP treatment independent of the particular physical regime
and provide a coherent state picture of electrons on the ambient
lattice, whatever the model interaction actually is. Even though
this approach is quite general, in the sequel we shall develop
it for the Hubbard Hamiltonian. 

It is well known that the Hubbard Hamiltonian can be rewritten in
terms of two coupled $XX$ models of $1/2$ spin operators by means of
the Jordan--Wigner transformation. Such a transformation can be
performed in any dimension as well as, in principle, for any electron
Hamiltonian, and leads quite naturally to a picture relying on spin
coherent states (SCS) \cite{ZFG}. When this is used explicitly
within the TDVP framework, the resulting equations of motion are
recognized to describe two coupled fluids, which dynamics we shall
discuss.

A basic trait of the spin description
is that its semiclassical version
is more reliable the more the spins are large \cite{LIE}.
Since this feature is in general not realized when starting from
quantum $1/2$ spin operators, we shall look here, in particular, for
solutions of the equations of motions corresponding to the macroscopic
excitations of few system modes, in which case we expect to describe
actual regimes for the Hubbard model itself. 
The problem of mode requantization, naturally in order due to the
expected quantum character of the low-temperature regime, is left
to a successive analysis \cite{BAS}.

The choice of $|\Psi \rangle$ as a direct product of single-site
Bloch states, representing the only assumption for our construction,
deserves some comments as to the expected reduction of the number of
states in the Hilbert space that are actually available for the system
dynamics. Such an effect usually occurs in a number of mean-field
approximations like the standard Hartree-Fock (HF) in which the
dominating features of the system are accounted for in an explicit
way thanks to an extreme reduction of the states accessible to the
system.

In this respect, using coherent states relative to the operators
of $H$ defined in the ambient lattice is by construction less
restrictive than using a subset of states tailored for a specific
regime. The advantage coming from this choice is manifold.
First, the structure of $|\Psi \rangle$ is however able to
produce an effective hamiltonian $\cal H$
that inherits both the nonlocal and the nonlinear character
of $H$, contrary to the Hartree--Fock (HF) scheme, in which
$H$ reduces to a sum of single-site linear Hamiltonians.
In passing, we notice that in many cases $\cal H$ exhibits a form
that is endowed with the same complexity of $H$.
In fact, the nontrivial form of $\cal H$ reflects the
basic character of the TDVP method that singles out $|\Psi \rangle$
variationally as the best solution to the original Schr\"odinger
equation \cite{ELBA}, whereas within the HF approximation what is
solved is a different Schr\"odinger equation, involving just the
linearized Hamiltonian. 

Second, as a consequence of the above feature, also the propagation
of any initial state is sustained by the full Hamiltonian, rather than
by its linearized HF version. Indeed, it is easily shown that the
latter entails quantum states whose time evolution is periodic,
while the TDVP dynamics is endowed with a much richer structure.
In particular, the dynamics of the expectation values of spin operators
(our dynamical variables) is consistently reproduced, whereas, when
turning to expectation values of products of spin operators, the
description obtained does not differ substantially from the one
that can be achieved within the random--phase approximation.

The Jordan--Wigner transformation
mentioned above amounts
to rewriting the electron annihilation operators $c_{j,\eta}$, with
$\eta =\uparrow,\downarrow$, in terms of Pauli spin matrices
$\sigma_{a,j}, \tau_{a,j}$, with $a=1,2,3$, which locally form
two (commuting) $su(2)$ algebras. For the Hubbard model, it turns
out that in dimension $D>1$ the possible transformed Hamiltonians
differ from each other due to a certain exponential factor in front
of the hopping term, which form in fact depends on the ordering
chosen for labeling the lattice sites. This problem has been already
investigated in the literature \cite{BAS}, and in the present paper
we shall limit our discussion to the one--dimensional (1D) case.
Explicitly, 
\be
c_{j,\uparrow} = P_j(\sigma_3)
\sigma_j^{-}\; , \quad
c_{j,\downarrow} =
P_L(\sigma_3)P_j(\tau_3) \, \tau_j^{-} \; ,
\label{JW}
\ee
where $P_j(\nu_3) \doteq \Pi_{\ell<j}\, \sigma_{3, \ell}$,
$\nu = \sigma, \tau$, from which the expressions for $\cjsd$
are straightforwardly derived. Here $L$ is the number of lattice
sites, $\nu^+_{j} \doteq \nu_{1,j} + i \nu_{2,j}$, with
$\nu = \sigma, \tau$. Remarkably, this transformation maps
fermions, which anticommute on different sites, into spins,
which commute on different sites, {\sl i.e.}
$[\sigma_{a,j},\sigma_{b, \ell}]= 0$
for $\ell \neq j$. 

Once Eqs. (\ref{JW}) are inserted into the Hubbard Hamiltonian,
the latter becomes
\be
H =  \sum_{j=1}^L [ U \sigma_{3,j}\,\tau_{3,j} -
T (\sigma^+_{j} \sigma^-_{j+1} + \tau^+_{j} \tau^-_{j+1}  + H.c.)] 
\; ,
\label{HUBSPIN}
\ee
when periodic boundary conditions are considered, and an odd
number of holes $N^h_\eta$ ($\eta =\uparrow, \downarrow$) on both
$\sigma$ sublattices is assumed \cite{SHA}, otherwise boundary
terms (corresponding to $j=L$) in the hopping contribution depending
on $T$ have to be rewritten as $ ({\rm e}^{\pi
(1 - N^h_\uparrow)}\sigma^+_{L} \sigma^-_{1} + {\rm e}^{\pi
(1 -N^h_\downarrow)}\tau^+_{L} \tau^-_{1}+ H.c.)$.
In Eq. (\ref{HUBSPIN})
the extra terms that take advantage of conserved quantities such as
the total electron number and the magnetization have been ignored.

In the next section, based on the spin--coherent--state picture, we
shall implement the TDVP procedure whereby one can derive from
(\ref{HUBSPIN}) the effective Hamiltonian and the related motion
equations.
In Sec. III, upon recognizing the two-fluid structure of the resulting
model, we shall solve explicitly the motion equations of each fluid
within a phase--locking approximation, and evidenciate how the Coulomb
interaction drives the system to a transition (apparently related
to the metal-insulator one) in which also the phases of the two
fluids become strongly locked. Tunneling phenomena between the
two fluids are also discussed.
In Sec. IV we specialize to the study of solutions exhibiting
a pure phase dynamics, and stress the aspect concerning the
macroscopicity of the excited degrees of freedom.
In Sec. V we show that the ground-state phase space known from
standard mean-field treatments can be obtained within our scheme by
analyzing the fixed points of a very simple collective phase solution,
corresponding, in fact, to describe the whole lattice as a sum of
two-site clusters. Finally Sec. VI is devoted to give some conclusions.

\section{coherent--states picture}

Approaching interacting spin systems within a semiclassical limit 
has been deeply investigated. In particular, it is well understood
that a consistent description can be obtained \cite{ZFG} by projecting
the Hamiltonian onto a basis of SCS.
In this case, an exact result obtained by Lieb \cite{LIE} shows that
the projected Hamiltonian reproduces the behavior of the original one
the more the spins are large, and in any case it gives upper and lower
bounds to the ground--state energy of the quantum Hamiltonian (the exact
value being recovered for infinitely large spins). 
One-half SCS are given by
\be
|\eta \rangle \equiv C(\eta) e^{\eta J_+ } |-1/2 \rangle \; ,
\label{COS}
\ee
where the maximum weight vector $|-1/2 \rangle$ belongs to
the $J_3$ spectrum ($J_3 |\pm 1/2 \rangle = \pm 1/2
|\pm 1/2 \rangle $) and fulfills the conditon
$J_- |0 \rangle =0 $, $J_-$ ($J_+ =(J_-)^+$)
representing the lowering (raising) operator. Also, defining the
normalization factor as $C(\eta) = 1/\sqrt{1+|\eta|^2}$ ensures the
condition $\langle \eta |\eta \rangle = 1$. The expectation values
of generators $J_3, J_{\pm}$
\be
S_3 = \langle J_3 \rangle = {\frac{|\eta|^2-1}{2(1+|\eta|^2)}}  
\ee  
\be
S_- = \langle J_- \rangle = {\frac{\eta}{(1+|\eta|^2)}}  
\ee
obtained by means of definition (\ref{COS}) ($\langle \bullet \rangle
\doteq \langle \eta| \bullet |\eta \rangle$), clearly exhibit their
semiclassical character when considering the fact that
$S_3$, $S_\pm$ satisfy the equation $S_3^2 + S_2^2 +S_1^2 =1/4$
(($S_+ \doteq S_1 + iS_2$)), namely the same sphere equation
fulfilled by the classic counterpart of the spin $(J_1, J_2, J_3)$
($J_+ \doteq J_1 + iJ_2$). In passing we notice that the spin
variables, assuming limited values, keep track of the fermionic
nature of the underlying system.

The set-up just developed can be readily extended to the
interacting spins of $H$. Assigning at each site a pair of
SCS $|\alpha_j \rangle $, $|\beta_j \rangle $ relative to the above
$\sigma$-spin and $\tau$-spin, respectively, allows one to implement
the TDVP procedure that is essentially based on constructing a
macroscopic trial wave function accounting for the microscopic
processes of the system. The simplest choice for a spin model
is realized through the state
\be
|\Psi \rangle \equiv e^{i S / \hbar}
|\alpha \rangle \otimes |\beta \rangle \; ,
\label{MWF}
\ee
where $|\alpha \rangle \otimes |\beta \rangle =
\otimes_j (|\alpha_j \rangle \otimes |\beta_j \rangle )$,
that provides the expectation values
$A^*_j =\langle \Psi| \sigma^{+}_j |\Psi \rangle$
($B^*_j =\langle \Psi| \tau^{+}_j |\Psi \rangle$)
and $A_{3j} =\langle \Psi|\sigma_{3,j} |\Psi \rangle$
($B_{3j} =\langle \Psi|\tau_{3,j} |\Psi \rangle$)
of $\sigma$ spins ($\tau$ spins). The description of the
microscopic dynamical activity in terms of such semiclassical
variables (actually they correspond to an ensemble of classical
spins) is achieved by showing that they obey a set of Hamiltonian
equations standardly derived from imposing $|\Psi \rangle$ to obey
the weaker version of the Schr\"odinger equation
$\langle \Psi|(i\hbar\partial_t - H )|\Psi \rangle =0$,
the latter requirement leading as well to interpret $S$ in
Eq. (\ref{MWF}) as the effective action. The explicit form of TDVP
Hamiltonian generating such hamiltonian equations turns out to be
$$
\langle H \rangle = \langle \beta |\otimes  \langle \alpha |
H |\alpha \rangle \otimes |\beta \rangle \ \; ,
$$
while the Poisson Brackets obeyed by the spin ensemble variables
implicitly follow from the equations of motion themselves.

Hubbard Hamiltonian (\ref{HUBSPIN}) in one-dimension,
when projected onto the trial state
$|\alpha \rangle \otimes |\beta \rangle $, becomes 
\bq
\langle H \rangle = N_s \frac{U}{4}  + \frac{U}{2} (A_{3} + B_{3})
+ U\sum_j A_{3j}\,B_{3j} + {\cal H}_T \; ,
\label{HAM}
\eq
where $A_{3} \doteq \sum_{j} A_{3j}$, $B_{3} \doteq \sum_{j} B_{3j}$
and the hopping term ${\cal H}_T$, which reads
\bq
{\cal H}_T \doteq -T \sum_{j} (A^*_{j} A_{j+1} +
B^*_{j} B_{j+1}+ H.C.) \; ,
\nonumber
\eq
is nothing but the sum of two classical XX models.
The Hamiltonian equations generated by the TDVP procedure
are given by 
\bq
i {\dot A}_{j} = (-\delta_A + UB_{3j}) A_{j} +
2T A_{3j} (A_{j_+1} + A_{j-1} ) \, ,
\label{SEA}
\eq
\bq
i {\dot B}_{j} = (-\delta_B + UA_{3j}) B_{j} +
2T B_{3j} (B_{j_+1} + B_{j-1} ) \, ,
\label{SEB}
\eq
\bq
i {\dot A}_{3j} = -T [ A^*_{j}\,( A_{j_+1} + A_{j-1}) -
A_{j}\,( A^*_{j_+1} + A^*_{j-1}) ] \, ,
\label{SEA3}
\eq
\bq
i {\dot B}_{3j} = -T [ B^*_{j}\,( B_{j_+1} + B_{j-1}) -
B_{j}\,( B^*_{j_+1} + B^*_{j-1})] \, ,
\label{SEB3}
\eq
where $\delta_A \doteq \mu_A - U/2$, $\delta_B \doteq \mu_B - U/2$,
once the Hamiltonian $\langle H \rangle$ is rewritten in the form
\be
{\cal H} \doteq \langle H \rangle + \mu_A \, \chi_A +\mu_B \, \chi_B 
\label{CONS}
\ee
containing the constraints $\chi_A \doteq \sigma_A -A_{3}$,
$\chi_B \doteq \sigma_B -B_{3}$ with Lagrange multipliers
$\mu_A$, $\mu_B$. The Poisson brackets implicitly entailed by
Eqs. (\ref{SEA})--(\ref{SEB3})
are given by
$$
\{C_j^* , C_j  \}= 2C_{3j}/i\hbar \, ,
\, \{C_{3j}, C_j^* \}= C_j^* /i\hbar
$$
with $C= A, B$, and exhibit the structure of a (classical) angular
momentum algebra. Also, they state that $A_{3}$, $B_{3}$, related
to the total number of spin-up and spin-down electrons by the formulas
$$
\langle \sum_j n_{j \uparrow} \rangle = A_3 + N_s /2 \; , \quad 
\langle \sum_j n_{j \downarrow} \rangle = B_3 + N_s /2 \; ,
$$
respectively, where $n_{j \sigma} = c^+_{j \sigma} c_{j \sigma}$,
($\sigma = \uparrow, \downarrow$),
are constants of motion since $\{A_{3}, {\cal H} \}= 0 =
\{B_{3} ,{\cal H} \}$.
It is thus natural investigating
spin dynamics when $A_{3}$, $B_{3}$ are assumed to have fixed values
$\nu_A \, , \nu_B $  by inserting such informations via the constraints
$\chi_A =0 = \chi_B$.  

The conservation, for each $j$,
of the Casimir functions $C_{Aj} = A^2_{3j} + |A_{j}|^2 $ 
and $C_{Bj} = B^2_{3j} + |B_{j}|^2 $ is preserved as well.
On the contrary, the total magnetization vector ${\bf M}=
(M_x, M_y , M_z) =\sum_j {\bf M }_j$ (where $M_x +i M_y \doteq M^+$
with
$M^+ = \sum_j \, \langle \Psi|  \sigma^+_j \tau^-_j |\Psi \rangle
= \sum_j A^*_{j} B_{j}$) is no longer conserved but only its
z component
$ M_z =\frac{1}{2} \sum_j \langle \Psi| (\sigma_{3,j} -\tau_{3,j})
|\Psi \rangle = \frac{1}{2} \sum_j (A_{3j} - B_{3j} )$.
In addition, we also notice that the usual particle-hole symmetry of
the quantum Hamiltonian survives at the semiclassical level, and it
is implemented by the particle-hole transformation $A_{3j}
\rightarrow -A_{3j}$ and $B_{3j}\rightarrow -B_{3j}$.
 
Two remarks are now in order. First, due to the choice of
macroscopic wave function (\ref{MWF}), Hamiltonian (\ref{HAM}),
and Eqs. (\ref{SEA})--(\ref{SEB3}) mantain the same structure of
Hamiltonian (\ref{HUBSPIN}) and of the ensuing Heisenberg equations
for the quantum spin variables, respectively,
which feature is nontrivial \cite{MOPE}.

Moreover, we notice that, when moving from the lattice description to
the continuum limit \cite{SC1} ($C_j \to C(x)=|C(x)| e^{i\theta (x)}$,
$x \in {\bf R}$, $C= A,B$), the resulting equations can be interpreted
as two nonlinear Schr\"odinger equations (NLSE) for the order parameter
fields $A(x)$, $B(x)$. A part from the nonlinearity issued from $C_{3j} =
\pm \sqrt{1/4 - |C_j|^2}$ that is capable of producing the standard
quartic term $|C_j|^4$ for $|C_j|^2 << 1/4$, a further contribution in
this sense comes from the Coulomb terms $U A_{3j}B_{3j}$. The standard
reduction of the nonlinear Schr\"odinger equation to the continuity
and the Bernoulli equation \cite{SC2} governing the dynamics of the
densitylike field $|C(x)|^2$ and the phase field $\theta (x)$,
respectively, suggests that Eqs. (\ref{SEA})--(\ref{SEB3}) can be
seen as describing the dynamics of a coupled two-fluid lattice
model.

\section{Two-fluid dynamics}
The two-fluid structure of Eqs. (\ref{SEA})--(\ref{SEB}) has been
recognized by reducing them to the standard form (cubic NLSE) thanks
to the assumption $|C_j|^2 << 1/4$, namely, considering low-density
fluids. In this regime the usual hydrodynamic picture is made far more
complicated by the presence of $A_{3j}$, $B_{3j}$ in front
of the off-site $T$ terms in Eq. (\ref{SEA}), and Eq. (\ref{SEB}).
In fact such factors, in addition to the usual Laplacian-like
terms of the (lattice) Schr\"odinger equation characterized by
$A_{3j},\, B_{3j} \simeq -1/2$, allow for the occurrence of
configurations where the $T$ terms exhibit anomalous signs
($A_{3j},\, B_{3j} >0$) through extended regions of the lattice.
The investigations of the corresponding dynamics is deferred to
a future study. 

A regime exhibiting, in a sense, an opposite character
($|C_j|^2 \simeq 1/4 \to C_{3j} \simeq 0 $) will be examined in
the present section. The two-fluid structure still characterizes
the motion equations even if the dynamics mainly concerns the phase
variables, the densitylike variables $|C_j|^2$ being now essentially
constant. It is worth noting as well how such a regime (characterized
by a Bernoulli-like dynamics) is nothing but that the quantum phase
regime naturally emerging from the XX model form of ${\cal H}_T$ 
for $|C_j|= const$. In fact, by setting first
\be
A_j =R_{j}\exp{i \alpha_j} \quad , \quad B_j =
S_{j}\exp{i \beta_j} \quad,
\label{MODFAS}
\ee
where $R^2_{j} \equiv 1/4 -A^2_{3j}$, $S^2_{j} \equiv 1/4 -B^2_{3j}$, 
consistently equipped with the standard canonical commutation relations
$\{\alpha_{\ell}, A_{3j} \} =
\delta_{\ell, j}/i \hbar =\{\beta_{\ell},B_{3j} \}$,
and recasting then Eqs. (\ref{SEA})--(\ref{SEB3})
in the action-angle variable version contained
in the Appendix, one is able to work out the two linear
second--order equations, 
\be
{\ddot \alpha_j} =4T^2 [w (\beta_{j+1}-2\beta_j+\beta_{j-1})
+(\alpha_{j+1}-2\alpha_j+\alpha_{j-1})] \, ,
\label{XYA}
\ee
\be
{\ddot \beta_j} = 4T^2 [w (\alpha_{j+1}-2\alpha_j+
\alpha_{j-1}) +(\beta_{j+1}-2\beta_j+\beta_{j-1})] \, ,
\label{XYB}
\ee
with $w = U/4T$, under the assumptions $|A_{3j}|, |B_{3j}|<<1/2$,
$(\alpha_{j+1}-\alpha_j) \approx 0 \approx (\beta_{j+1}-\beta_j)$.
Eqs. (\ref{XYA})--(\ref{XYB}) describe dynamics of first order
quantities and exhibit the lagrangian structure typical of two
classical planar XX models nontrivially
phase coupled for any nonvanishing $U \ne 0$.

Remarkably Eqs. (\ref{XYA}) and (\ref{XYB}) can be decoupled (and
solved) upon defining $\theta_j= \alpha_j+\beta_j$, $\varphi_j=
\alpha_j-\beta_j$. In this case they become
\bq
{\ddot \theta_j} &=& 4T^2 (1+ w) (\theta_{j+1}-2\theta_j+\theta_{j-1}) \;,
\nonumber \\
{\ddot \varphi_j} &=& 4T^2 (1-w)
(\varphi_{j+1}-2\varphi_j+\varphi_{j-1}) \; , \label{XYP}
\eq
whose solution can be easily worked out in terms of Fourier modes.
In particular, let us notice that the parameter $w$ plays a relevant
role, in that it drives the $\varphi$ dynamics of the system from an
oscillatory regime ($w<1$) to a damped one ($w>1$), whereas the
$\theta$ dynamics remains purely oscillatory.
This is explicit when considering any single mode solution
of the form $\varphi_j(t;q) = \cos( \lambda_q t + \nu_j)$
and the ensuing dispersion relation
\be
\lambda^2_q = 16 T^2 (1-w)\sin^2( \pi q/L) \;.
\label{DISREL}
\ee
In terms of the original
phases $\alpha_j$ and $\beta_j$ this implies a phase-locking
phenomenon for $w>1$ ($U>4T$), which is physically quite
natural the more the on-site Coulomb repulsion becomes large.
Having in mind the metal-insulator transition typical of the Hubbard
model, which takes place at analogous values of $U$, we can argue
that the change in the dynamical behavior parametrized by $w$
might bear memory of such transition.  

It is worth noting that, again to the first order, Eqs. (\ref{SEA3})
and (\ref{SEB3}) for $A_{3j}$, $B_{3j}$ reduce to
\be
{\dot A}_{3j} = -(T/2)(\alpha_{j+1}-2\alpha_j+\alpha_{j-1}) \, ,
\label{EA3}
\ee
\be
{\dot B}_{3j} = -(T/2)(\beta_{j+1}-2\beta_j+\beta_{j-1})
\label{EB3}
\ee
which, despite the approximation introduced, still shows a nontrivial
time dependence of $A_{3j} $, $B_{3j}$.
The comparison of the above equations with those describing 
the tunneling phenomena of Josephson junctions \cite{TITI} 
is quite natural, coming from the fact the same equations can
be obtained, in the same linearized form, when considering the 
Josephson-junction array Hamiltonian that can be represented in
the simplified form by 
$H_{JJ} = \sum_j C^2_{3j} - g \sum_j cos(\gamma_{j+1} -\gamma_j)$
\cite{JOJU}.
This is confirmed as well by Eqs. (\ref{EA2}) and (\ref{EB2}) of
the Appendix which, within the present approximation
($R_j, S_j \simeq 1/2$), reproduce exactly the equation
${\dot C}_{3j} =\{ C_{3j},H_{JJ}\}$ for the on-site charges $C_{3j}$.
The special trait characterizing $\cal H$ is the quadratic term
$A_{3j} B_{3j}$ that generates a coupled phase dynamics via
Eqs. (\ref{XYA}) and (\ref{XYB}), namely a linearized system of two
$U$-coupled arrays. Also, this suggests to define here a quantity
that describes the net local current between the two arrays.
If we let $A_j$ and $B_j$ play the role of the  
Josephson wave functions, and $A_{3j}$, $B_{3j}$
as on-site charges, such current turns out to satisfy
the equation 
\be
I_j \simeq -{T\over 2}
(\varphi_{j+1}-2\varphi_j+ \varphi_{j-1}) \, ,
\ee
where $I_j \doteq \dot A_{3j}-\dot B_{3j} $. Hence the tunneling
phenomenon keeps track itself of the dependence on $w$,
vanishing in the strong Coulomb repulsion regime ($U> 4T$).

\section{phase dynamics}
Apart from the case related to Eqs. (\ref{EA3}) and (\ref{EB3}),
in the present paper we shall investigate solutions of Eqs.
(\ref{SEA})--(\ref{SEB3}) such that only the phases play a relevant
dynamical role, $A_{3j}$ and $B_{3j}$ being constant in time.
If, on the one hand, the dynamical situations in which
$A_{3j}$, $B_{3j}$ are involved exhibit a complex behavior and
their investigation goes beyond the purposes of the present paper,
on the other hand, considering only $\alpha_j$, $\beta_j$ as
dynamically active still entails situations that are far from
being trivial and facilitates the recognition of the topological
features that possibly characterize the solutions.

Hamiltonian (\ref{HAM}) describes the dynamics of interacting
classical angular momenta.  The latter exhibits solutions that
consistently match the semiclassical nature of the present
approach the more, by appropriately changing the basis of canonical
coordinates, one identifies some new variables
that could assume macroscopically large values and exhaustively
account for the system dynamics \cite{MAT}. 
In general, for a given dynamical  system, the excitations
corresponding to the proper dynamical modes (if any) provide
both the simplest and natural way to construct macroscopic
semiclassical solutions.  Unfortunately the identification of
proper modes is equivalent to making explicit solution of the
Hamiltonian equations, which in our case are highly nonlinear.
Nevertheless, based on the usual Fourier modes picture, where
$$
C_j=L^{-1/2} \sum_{k=1}^L {\exp}{ (i \tilde k j)}
\tilde C_k \; ,
$$
with $\tilde k = 2 \pi k /L$, $C = A, B$, one may wonder
whether there exists any integrable case corresponding to associate
the macroscopically large number of spin degrees of freedom
with a finite number of excited Fourier modes. It turns out that
this is the case at least for two classes of solutions.

\subsection{Vortex dynamics}
First, it is easily verified that the case corresponding to two
single excited Fourier modes $p$ and $q$, one for each fluid,
i.e., $\tilde A_p\doteq L^{1/ 2} R_A$,  $\tilde A_k= 0,
k\neq p$, and $\tilde B_q \doteq L^{1/ 2} R_B$, $\tilde B_k=0,
k\neq q$, is solution of Eqs. (\ref{SEA})--(\ref{SEB3}) with
\bq 
A_j (t) &=&
R_A {\rm exp} \{ i [(j \tilde p-
\omega_{A}{(p)} t + \phi_A)] \} \, ,
\label{PRA} \\ 
B_j (t) &=&
R_B {\rm exp} \{i [j \tilde q-
\omega_{B}{(q)} t + \phi_B)] \} \, ,
\label{PRB}
\eq
where $R_C \equiv {\sqrt{{1\over 4}-C^2}}$ with $C = A, B$,
and $A=A_3/L$, $B=B_3/L$, $\phi_A, \, \phi_B$ are arbitrary
phases accounting for the $U(1)$ symmetry of dynamical equations
and
\be
\omega_A {(p)}=(-\delta_A +U B)+ 4 T A \cos \tilde p \; ,
\label{FA}
\ee
\be
\omega_B{(q)} = (-\delta_B + U A)+4 T B \cos \tilde q \; .
\label{FB}
\ee
The corresponding energy per site is 
straightforwardly obtained as
\be
E_{p,q}= U(A+ 1/2)(B+1/2) -
2T [R^2_A \cos\tilde p+ \! R^2_B \cos\tilde q] \; . 
\label{TOP}
\ee
The main feature of solutions (\ref{PRA}) and (\ref{PRB}) is their
topological character encoded by the winding  numbers $p$ and $q$.
Notice that we have assumed periodic boundary conditions providing
our 1D lattice with the topology of the circle, and $A_j$, $B_j$
can be regarded as order parameters covering two ${\bf S}^1$
configuration spaces. 
Within this picture the indices $p$ and $q$ account for the number
of times  $A_j$ and $B_j$ cover their configurations spaces while
$j$ goes from $0$ to $L$.  Indeed such configurations are nothing but 
1D vortex excitations once the phases of the order parameters are
identified with the potential functions of two coupled fluids. Here
the coupling is fully contained in the frequencies $\omega_{A}(p)$
and $\omega_{B}(q)$.

Interestingly, it is possible to evaluate explicitly correlation
functions for solutions (\ref{PRA}) and (\ref{PRB}). Their physical
meaning is better understood when writing them for the original
fermionic system. In this case, two-site correlations within a single
fluid (e. g., the one with up spins), read 
\be
\langle c_{j\uparrow}^\dagger c_{\ell\uparrow}+ H.c. \rangle 
=2 (2 A)^{|\ell - j|-1}
R_A \cos [\tilde p (j-\ell)] \, , \; 
\ell\neq j \quad ,
\ee
whereas for sites belonging to the two different fluids are 
$$
\langle c_{j\uparrow}^\dagger c_{\ell\downarrow} + H.c.\rangle =
2 (2 A)^{L-j} (2 B)^{\ell -1} R_A R_B 
$$
\be
\times \cos \{ j \tilde p-\ell \tilde q+
[\omega_B(q)-\omega_A(p)] t + (\phi_A -\phi_B) \} \,.
\ee
with $j \neq \ell $. As expected, in both cases
long-range order does 
not emerge since $2|A|$, $2|B|$ are smaller than one in any nontrivial
case. However, two remarkable features emerge. First, they manifestly
keep track of the topological character of the solution through the
winding numbers $p$ and $q$. Second, but more important, the two-fluid
correlation function also exhibits a time-dependent behavior, whenever
the density of the two fluids or the topological charges are different.
This last feature should be viable to experimental observation.

\subsection{Staggered dynamics}
The general class of solutions characterized by the phase dynamics
is obtained when $B_{3j}, \, A_{3j}$ are assumed to be assigned. In
this case Eqs. (\ref{SEA})--(\ref{SEB3}) reduce to a linear system of
equations for the variables $A_{j}$'s, and $\, B_{j}$'s where proper
modes coincide with the eigenvalues of a certain secular equation.
In fact, one should recall that assigning $B_{3j}$, $A_{3j}$ and
thereby reconstructing $|B_{j}|$, $|A_{j}|$, leaves the possibility
to satisfy the eigenvalue problem by exploiting just the phases of
$B_{j}$ and $A_{j}$.

For $A_{3j}$  and $B_{3j}$ constant in time,
Eqs. (\ref{SEA3}) and (\ref{SEB3}) are conveniently rewritten
(see the Appendix) in terms of
action-angle-like variables defined in (\ref{MODFAS}), as 
\be
R_{j+1} \sin(\alpha_{j+1}-\alpha_j ) +
R_{j-1} \sin(\alpha_{j-1}-\alpha_j ) =0 \; ,
\label{PD1}
\ee
\be
S_{j+1} \sin(\beta_{j+1}-\beta_j ) +
S_{j-1} \sin(\beta_{j-1}-\beta_j ) =0 \; .
\label{PD2}
\ee
The general solution is not known. Of course a simple solvable
case is obtained by assuming both $R_{j}$ and $S_{j}$ constant and
independent of $j$. This leads to the vortex case discussed
in the previous subsection.
A further solution exhibiting an interesting dynamics is obtained
by noticing that $R_{j+1}$, $R_{j-1}$, can be factored out from the
above conditions upon assuming that $R_{2\ell} = R_E$ and
$R_{2\ell+1} = R_O$, $\forall \ell$, with $R_E$, $R_O$ fixed
constants.
The same assumptions can be implemented on $S_{j+1}$, $S_{j-1}$,
so that when they are inserted in Eqs. (\ref{PD1}) and (\ref{PD2}),
these turn out to depend only on the difference
$\gamma_{j+1}-\gamma_j$, with $\gamma = \alpha, \beta$.
The latter has two possible values satisfying the equations,
$\gamma_*$ or $\pi-\gamma_*$ for each $j$, with $\gamma_*$
time-dependent function.
Then Eqs. (\ref{SEA}) and (\ref{SEB}) can be solved explicitly,
when rewriting them in the action-angle form of the Appendix.
In fact, it turns out that a consistent solution is achieved provided
$\gamma_{2j+1}-\gamma_{2j} \equiv \gamma_*$, and
$\gamma_{2j}-\gamma_{2j-1}=\pi -\gamma_*$,
for each $j$, which entails
\be
\gamma_{2j+1} =
j \pi + \gamma_1\;,\; \gamma_{2j} = (j-1) \pi + \gamma_2 \;.
\label{SEO}
\ee
$\gamma_1$ and $\gamma_2$ are time--dependent functions responsible
for the system's phase dynamics as solutions of the corresponding
equations given in Eqs. (\ref{EA1}), (\ref{EB1}). For instance
in the case $\gamma=\alpha$ they read
\bq
\alpha_1 &=& (\delta_A-UB_{3O}) t + \alpha_1(0) \;,\; \nonumber \\
\alpha_2 &=& (\delta_A-UB_{3E}) t + \alpha_2(0) \; ,
\label{SEO2}
\eq
while the analogue for $\beta_1$, $\beta_2$ is easily derived.
Interestingly, the time-dependent part of the phases keeps
track of the coupling between the two fluids for any nonvanishing
value of the Coulomb repulsion $U$. Again, such a feature
should be viable for experimental observation.

Apart from the initial conditions $\gamma_1(0)$, $\gamma_2(0)$,
the solution (\ref{SEO}), (\ref{SEO2}) clearly exhibits a
staggering in the phases both on the even and on the
odd sublattices. Making such a solution consistent with
periodic boundary conditions constrains the length of the lattice
$L$ to be $L=4 p$, $p\in {\bf N}$. Once more this feature can be
related to the macroscopic excitation of some Fourier modes
(two for each fluid). Explicitly for $C=A$
\bq
A_{L/4}&=& {1\over 2}\sqrt{L}
\left [R_E {\rm e}^{i \alpha_2(0)}+ i R_O
{\rm e}^{i  \alpha_1(0)}\right] \; , \nonumber\\
A_{3L/4} &=& {1\over 2}\sqrt{L}
\left[R_E {\rm e}^{i \alpha_2(0)} - i R_O{\rm e}^{i  \alpha_1(0)}
\right] \; ,  \nonumber\\
\label{SEO3}
\eq
and $A_k = 0$ for $k\neq p, 3 p$, the analogue holding as well for
$C= B$, $\phi=\beta$.

The minimum energy per site $E_s$ of the above staggered solution
--to be compared with successive results for different phases--
is found to be $E_s = U(\nu+|\nu|)/4 $. It is important to observe
how the independence of $E_s$ from $T$ (to be interpreted as the
absence of a net global current) follows from the fact that 
the contributions to the hopping term coming from subsequent
lattice bonds, let us say ($j, j+1$) and ($j+1, j+2$), cancel
each other.
At the microscopic level, however, the hopping terms actually
contribute in terms of local currents
(these are essentially given by
$\gamma_{2j+1}-\gamma_{2j} = \gamma_*$,
$\gamma_{2j}-\gamma_{2j-1} = \pi -\gamma_*$)
with opposite sign.

\subsection{Many-sublattices solution}

Further solutions to Eqs. (\ref{SEA})--(\ref{SEB3})
that corresponds to the excitation of a finite number of Fourier
modes (endowed with a macroscopic character) can be recovered by
partitioning first the lattice $\Lambda$ into $n = L/q$ sublattices
$\Lambda_a$ of $q$ sites ($q$ divisor of $L$), and introducing then
the collective variables
\be
A_a \doteq \sum^{q-1}_{\ell = 0} \, A_{\ell n +a} \, ,\;
A_{3a} \doteq \sum^{q-1}_{\ell = 0} \, A_{3(\ell n +a)} \, ,
\label{collec}
\ee
with $\ell \in (0, q-1)$, $a \in (1, n)$.
Here $A_a$, $A^*_a$, and $A_{3a}$ still
fulfill the commutation relations of a (classical) algebra $su(2)$.
It turns out that Eqs. (\ref{SEA})-(\ref{SEB3}) can be
rewritten in terms of the above collective variables provided 
further assumptions are stated. These are $A_{3j} = A_{3a}/q$,
$A_j = A_a/q$ with $j \in \Lambda_a$. When this is the case,
dynamical equations reduce to a set of $4L/q$ equations now
written in terms of $ A_{3a}$, $A_a$, $B_{3a}$, $B_a$ exhibiting
the same structure. In the Fourier transformed space this amounts
to the excitations of $n$ modes, i.e.,
\be
\tilde A_k = {1\over\sqrt{N}} \sum_{a=1}^{n}
{\it e}^{i \tilde k a} A_a \, , 
\quad \tilde A_\ell=0 \, , 
\ee
($\tilde k = 2\pi k/N$) for $k= m q$, $\ell \neq m q$ ($0<m \leq n $),
respectively. Solutions within this class are now obtained by solving
the remaining $4n$ equations, which preserve the same complex
structure of the original ones.

For the simplest case $n=2$ ($n=1$ being a subclass of
vortexlike solutions) the dynamical equations are represented by
\be
i {\dot A}_{1} = (-\delta_A + UB_{31}) A_{1} +4T A_{31} A_{2} 
\label{TSA1}
\ee
\be
i {\dot B}_{1} = (-\delta_B + UA_{31}) B_{1} + 4T B_{31} B_{2}  
\label{TSB1}
\ee
\be
i {\dot A}_{2} = (-\delta_A + UB_{32}) A_{2} +4T A_{32} A_{1} 
\label{TSA2}
\ee
\be
i {\dot B}_{2} = (-\delta_B + UA_{32}) B_{2} + 4T B_{32} B_{1} \, ,
\label{TSB2}
\ee
together with those for $A^*_{j}$ and $B^*_{j}$.
Correspondingly Hamiltonian (\ref{HAM}) takes the form
$$
H_{2} = \frac{N_s}{2} \{ U/2 - \sum_{C=A,B}\,
[\delta_C (C_{31} +C_{32}) -\nu_C ]+
$$
\bq
U (A_{31}B_{31} + A_{32} B_{32}) 
-2T(A_{1}A^*_{2} + B_{1}B^*_{2} + C.c.) \} \, .
\label{HAM2}
\eq
As the number of first integrals of motions is 3 ($H$, $A_3$, and
$B_3$), whereas the equations are now 8, this case is
nonintegrable. However, being interested in phase dynamics in which
case $A_{3j}$ and $B_{3j}$ are constants for each $j$, the solution
to Eqs. (\ref{TSA1})-(\ref{TSB2}) can be worked out explicitly.
The latter is characterized by collective frequencies $\lambda_A$,
$\lambda_B$ for the $A_j$'s and $B_j$'s of the form $C_j =
C_j(0) \exp (i\lambda_C t)$ ($C= A, B$, $j= 1,2$) which are
independent from each other.

It is important to notice how the case presently studied differs
from the staggered solutions described above since $C_{j+2} = C_j$
is not contained in Eqs. (\ref{PD1})--(\ref{PD2}), namely
${\rm Im} [C^*_{j}( C_{j+1} + C_{j-1})]=0$. When $C_j(t)$ are
inserted in Eqs. (\ref{TSA1})--(\ref{TSB2}) one is able to recast
them in the form
\bq
U(B_{31} -B_{32}) =
4T \left (A_{32} \frac{A_1}{A_2} - A_{31} \frac{A_2}{A_1} \right) \, ,
\label{FP1}
\eq
\bq
U(A_{31} -A_{32}) =
4T \left (B_{32} \frac{B_1}{B_2} - B_{31} \frac{B_2}{B_1} \right) \, ,
\label{FP2}
\eq
\bq
2 \delta_A -\lambda_A = U \nu_B +
4T \left (A_{32} \frac{A_1}{A_2} + A_{31} \frac{A_2}{A_1} \right) \, ,
\label{FP3}
\eq
\bq
2 \delta_B -\lambda_B = U \nu_A +
4T \left (B_{32} \frac{B_1}{B_2} + B_{31} \frac{B_2}{B_1} \right) \, ,
\label{FP4}
\eq
where $C_j$ ($C_{3j}$) stay for initial conditions $C_j(0)$
([$C_{3j}(0)$], and the constant of motion
\be
\nu_A \equiv A_{31} + A_{32}\, , \,\nu_B \equiv B_{31}+B_{32}
\label{FILS}
\ee
are input data, whereas $\delta_A$, $\delta_B$, $A_{3j}$, $B_{3j}$
(consistently with $\nu_A, \nu_B =$ const) are the unknown parameters
to be fixed. 

It is worth noticing that Eqs. (\ref{FP1}) and (\ref{FP2}) turn
out to be completely independent from $\lambda_A$, $\lambda_B$
while in Eqs. (\ref{FP3}) and (\ref{FP4})
$\lambda_A$ and $\lambda_B$ can be
incorporated inside $\delta_A$ and $\delta_B$ by redefining them
as $\Delta_C = \delta_C - \lambda_C/2 $, $C=A,B$. At the operative
level this fact allows one to reconstruct the solution of 
(\ref{FP1})-(\ref{FP4}) for $\lambda_C \ne 0$ from the case
$\lambda_C =0$, which by the way identifies the fixed points of
Eqs. (\ref{TSA1})-(\ref{TSB2}). The investigation of such points
is deepened in the next section. 

\section{Fixed points of two-sublattice solution}

The present approach is able to give a (simplified) description
of the system dynamics, with a number of interesting features, as
we have seen in the previous section. Nevertheless, as a secondary
effect, it also gives the system equilibrium states, which coincide
in fact with fixed points of the equations of motion. Many other
(mean-field-like) approaches are focused on the study of equilibrium
and especially ground states of Hamiltonian (\ref{HAM}).
For instance, from the Hartee-Fock approximation \cite{HIR} it is
known that the $T=0$ phase space contains an antiferromagnetic, a
ferromagnetic, and a paramagnetic phase for $U>0$. In this section we
shall see that a similar description of the ground-state phase can be
already obtained by studying fixed points of the simple two-sublattice
solution, the latter being given by Eqs. (\ref{FP1})-(\ref{FP4}) for
$\lambda_C=0$.
	
In particular, as Eqs. (\ref{FP3}) and (\ref{FP4}) just fix
$\delta_A$, $\delta_B$, we search for the solutions of
Eqs. (\ref{FP1}) and (\ref{FP2}), in which the unknowns are two.
It is convenient to introduce the pair of
new variables $a=A_{31} A_{32}$, $b=B_{31} B_{32}$, in terms of which
the above equations reduce to the pair of fourth order, coupled
equations
\bq
(\nu_B^2-4b)=
g^2 \frac{(1 + 4a)^2 (\nu_A^2-4 a)}
{(1 + 4a)^2 -4\nu_A^2} \,
\label{EQ1}
\eq
\bq
(\nu_A^2 - 4 a) =
g^2 \frac{(1 + 4b)^2 (\nu_B^2-4 b)}
{(1 + 4b)^2 -4\nu_B^2} \; ,
\label{EQ2}
\eq
with $g = 4 T/U$. The two equations (\ref{EQ1}) and (\ref{EQ2}) can
be recast into a single eight-degree equation for the variable $a$,
$$
[Z^2 (1-g^4) -4\nu_A^2]
[ (1+\nu_B^2)(Z^2 - 4\nu_A^2)-
$$
\be
g^2 Z^2 (\nu_A^2 +1-Z) ]^2 =
4 \nu_B^2 (Z^2 - 4\nu_A^2)^3 \, ,
\label{EQ3}
\ee
where $Z \doteq 1 +4a $, and
the factor $\nu_A^2- 4a$, which provides an independent solution,
has been factored out [see (i) below]. The variable $b$ is then
easily worked out from Eq. (\ref{EQ1}).

First, let us notice that the independent solution
$\nu_A^2-4a =0$ implies $\nu_B^2-4b =0$ leading to ${a}= \nu_A^2 /4$,
${b}= \nu_B^2 /4$. As $\nu=\nu_A+\nu_B =2(s_a \sqrt{a} +s_b \sqrt{b})$
with $s_a =\pm 1$, $s_b =\pm 1$, this solution implies $A_{31}=A_{32}
= \nu_A/2$, $B_{31}=B_{32}= \nu_B/2$ (two-sublattice solutions with
ferromagneticlike order on each sublattice). It has energy 
\be
2 \frac{H}{N_s} = \frac{U}{2} \left [1+ \nu +\nu_A \nu_B +
\frac{g}{2} (\nu_A^2 + \nu_B^2-2) \right] \; ,
\label{equal}
\ee
which matches the one of vortex solution $E_{p,q}$
[see Eq. (\ref{TOP})] in the untwisted case $p=q=0$.
For fixed filling $\nu=\nu_A+\nu_B$ its minimum value depends on the
actual value of $p$. When $g <1$ (i.e., $U>4$, this is reached
either for $\nu_A=\nu\, , \nu_B=0$ or for $\nu_B=\nu\, ,\nu_A=0$,
in which case the energy becomes
\be
E_{pf}=\frac{U}{2}
\left (1+\nu - g +\frac{g}{2} \nu^2 \right) \; ,
\label{eferro}
\ee
the solution describing ferromagnetism away from half-filling within
a single cluster, in that the average magnetization on the cluster
$M=(\nu_A-\nu_B)/2$ coincides with $\pm\nu/4$. The subindex $p$ in
$E_{pf}$ is to remind us that the solution on the lattice, due to the
arbitrary choice of the sign of $M$ on each cluster, does {\it not}
exhibit ferromagnetic order.

On the contrary, for $g>1$ the minimum value of expression
(\ref{equal}) is reached when $\nu_A=\nu_B=\nu/2$. Physically, it
corresponds to a paramagnetic solution even within a single cluster,
and has energy
\be
E_p =\frac{U}{2} \left [(1+ {\nu\over{2}})^2 -
g (1-{\nu^2\over 4})\right]
\, . \label{epara}
\ee

Having in mind the phase diagram known from mean-field-like
Hartree-Fock treatment of the Hubbard model, an antiferromagnetic
solution is also expected, where the energy should be lower than both
$E_f$ and $E_p$ near half-filling and for large $U$. 
This can be worked out as a solution of Eq. (\ref{EQ3}) when the
magnetization is zero, namely $\nu_A=\nu_B$. In this case, it is
easily realized that Eq. (\ref{EQ3}) can be rewritten as the product
of a second-order factor $(g^2 -1)Z^2 + 4\nu^2 $ (real for  $g < 1$)
and a sixth-order factor which, in the range of parameters physically
allowed, never provides real solutions.  
On the contrary, the vanishing of the second-order factor
in $a$, in fact, leads to an antiferromagnetic solution. This
can be seen by first realizing that an analogous equation holds
also for $b$, so that finally Eq. (\ref{EQ1}) and (\ref{EQ2}) reduce
to two second-order ones
\be
1={\frac{ g^2 (1 + 4a)^2}{(1 + 4a)^2 -4 \nu_A^2}} \, ,
\label{RA}
\ee
\be
1={\frac{g^2 (1 + 4b)^2}{(1 + 4b)^2 -4 \nu_B^2}} \, ,
\label{RB}
\ee
which in order to consistently match $\nu_A \equiv \nu_B$
imply $a=b$, with
\bq
b= a = \frac{1}{4} \left ( \frac{2| \nu_A|}{\sqrt{1-g^2}} -1 \right)
\, .
\label{FB}
\eq
Notice that, away from half-filling (which corresponds to
$\nu_A=\nu_B=0$), the condition $g < 1$ follows from
Eqs. (\ref{RA}) and (\ref{RB}). Moreover, when
calculating explicitly $A_{31}$ ($A_{32}=\nu_A-A_{32}$) and $B_{31}$
($B_{32}=\nu_B-B_{32}$) through formula (\ref{FB}) which read
\be
A_{31} \equiv \frac{1}{2} \left ( \nu_A \pm \sqrt{\nu^2_A +1  
-2 \frac{|\nu_A|}{\sqrt{1-g^2}} } \right) \; ,
\label{GA}
\ee
\be
B_{31} \equiv \frac{1}{2} \left ( \nu_B \pm \sqrt{\nu^2_B +1  
-2 \frac{|\nu_B|}{\sqrt{1-g^2}} } \right) \; ,
\label{GB}
\ee
one singles out the further restriction
$|\nu_A|=|\nu_B| < \sqrt{(1-g)/(1+ g)}$.
The apparent freedom in choosing the sign in
Eqs. (\ref{GA}) and (\ref{GB})
just corresponds to exchange the role of $A_{31}$ and $A_{32}$
($B_{31}$ and $B_{32}$).
In fact, the physical solutions turn out to be just two,
in that the condition  
$a=b$ can be implemented in two different ways, namely
$A_{31}=B_{31}$, $A_{32}=B_{32}$ (paramagnetic), and
$A_{31}=B_{32}$, $A_{32}=B_{31}$ (antiferromagnetic). 
The energies corresponding to such solutions, 
$$
E_{p}'= \frac{U}{2}( 1+ \nu) + U(A_{31}^2 + A_{32}^2) -
8T \sqrt{\frac{1}{4}-A^2_{31}}  \sqrt{\frac{1}{4}-A^2_{32}} \; ,
$$
$$
E_{af}= \frac{U}{2}( 1+ \nu) + 2 U A_{31} A_{32} -
8T \sqrt{\frac{1}{4}-A^2_{31}}  \sqrt{\frac{1}{4}-A^2_{32}} \; , 
$$
differ only due to the $U$ term, which is manifestly lower in the
antiferromagnetic case. 
It turns out that the antiferromagnetic cluster energy
\bq
E_{af} = \frac{U}{2}( \nu + |\nu| \sqrt{1-g^2}) \label{eaf}
\label{ENER3}
\eq
is always lower than paramagnetic case within the domain specified
by $|\nu_A|=|\nu_B| < \sqrt{(1-g)/(1+g)}$.
 
The successive comparison among Eqs. (\ref{epara}),
(\ref{eferro}), and (\ref{eaf}) shows that the
ground-state phase space for this
two-sublattices solution (Fig. 1) exhibits a structure in 
qualitative agreement with many other theories, in particular the one
obtained in the low-density approximation for the one--band model
(Ref. \cite{MAT}).
Moving from half-filling, in which case a magnetic phase is
found for $U>4T$, the antiferromagnetic phase takes place at increasing
values of $U$, and in any case for filling greater than one quarter.
Indeed, by requiring that $E_{af}< E_{pf}$, the transition line to
the antiferromagnetic phase is given by
\be
\nu=\frac{1}{g} \left (g-1+\sqrt{1-g^2}\right )>0 \, .
\ee
For lower values of filling, the system is a nonmagnetic metal.
Within such regime an extra transition emerges for $g=1$ from a
paramagnetic solution with ferromagnetic structure on each cluster
(energy $E_{pf}$), and a paramagnetic solution with no order even
within the clusters (energy $E_p$). Apparently by increasing $U$ the
lattice begins to organize towards ferromagnetism. 
Let us notice that, consistently with the 1D character of the model
studied, both ferromagnetic and antiferromagnetic solutions exhibit
only local order, in that the actual value of the magnetization on
different two-site clusters is uncorrelated.

In the previous section we explicitly gave the energies corresponding
to some simple solutions of the equation of motions exhibiting
nontrivial dynamics. A natural question is then whether some of these
solutions survive down to the ground state, or not. Interestingly,
one can verify that, in fact, the staggered solution, with energy
$E_s$, at half-filling turns out to be degenerate with the two-site
antiferromagnetic solution described above, with energy $E_{af}$.
Indeed, both of them in this case have a vanishing hopping term,
in agreement with the expected insulating behavior of such regime,
and in practice on the single cluster the two solutions coincide.
However, the explicit solution of the equations of motion in the
staggered case proves that at a dynamical level the only consistent
way of moving from the fixed point is by means of the staggered
choice of phases.
\section{conclusions}
The main object of this paper has been to develop
an approach to the Hubbard model quantum dynamics that is not
based on the particular physical regime under investigation, on the
one hand, and is capable of reformulating the model dynamics in a
form more tractable than that relying on the direct diagonalization
of the model Hamiltonian, on the other. Such requirements
have been achieved by combining three ingredients, which are the
representation of quantum dynamics within a coherent-state picture,
the expression of the Hubbard Hamiltonian in terms of spin variable
(\ref{HUBSPIN}) issued from its fermionic standard form through
the Jordan-Wigner transformation (\ref{JW}),
and the implementation of the TDVP method.
The choice of the trial state (\ref{MWF}) has generated Hamiltonian
(\ref{HAM}) (that is $\cal H$ with the constraints $\chi_C =0$, $C=
A,\, B$) whose dynamics is governed by Eqs. (\ref{SEA})--(\ref{SEB3}),
and accounts for the evolution of the spin operator expectation
values.

The resulting dynamical scenery has revealed both a rich
structure --that corresponding to a pair of XX models coupled through
the Coulomb term-- and interesting links with other models.

For $|A_j|^2,\, |B_j|^2 \simeq 0$, one obtains a model of two coupled
fluids at low density. In particular, in this limit, the dynamics has
been recognized to have the form of two coupled lattice NLSE.
A feature that is unusual for the standard NLSE comes from the
dependence of the off-site terms in Eqs. (\ref{SEA}) and (\ref{SEB})
on the signs of $C_{3j}$, which allows for the fragmentation of the
planar lattice in regions where either $C_{3j} \simeq +1/2$
(sites occupied by electrons of type $C$), or $C_{3j} \simeq -1/2$
(local depletion of electrons of type $C$). The latter case suggests
the occurrence of soliton-like behavior in correspondence to the
negative sign of the off-site terms. 

The opposite regime $|A_j|^2,\, |B_j|^2 \simeq 1/4$
has been studied in Sec. III, where the equations
of motion have revealed that the model actually
describes two coupled Josephson-junction arrays.
In particular the condition $|C_j|^2 \simeq 1/4$ makes emerge
a dynamics concerning essentially the phases $\alpha_j$, $\beta_j$
that can be solved exactly after reducing the equations of
the Appendix to the linear system described by Eqs. (\ref{XYA}) and
(\ref{XYB}).
Its main feature is certainly the macroscopic effect
of phase locking [$(\alpha_j -\beta_j) \to 0$] which is
enacted when going from $U< 4T$ to $U> 4T$, and might be related
to the metal-insulator transition exhibited by the Hubbard model.

Pursuing the investigation of dynamical situations
in which phases are active and $|A_j|^2, |B_j|^2 = const$ has
led us to recognize two other interesting results.
First, a set of topological solutions has been obtained by
considering uniform configurations $C_{3j} = C_3 /L$, $C= A, B$,
which are nontrivial when one excludes the half-filling case.
The phases $\alpha_j$ and $\beta_j$ are allowed to change as $j$ is
varied so as to give rise to a pair of vortexlike configurations
labeled by two integers $p$ and $q$ for $A_j$ and $B_j$ (the fluid
order parameters), respectively. Also, the time behavior exhibits a
dependence on the electronic fillings as well as on the topological
characters through the frequencies $\omega_A (p)$ and $\omega_B (q)$.

A second class of solutions has been obtained instead when considering
the solutions of Eqs. (\ref{SEA}) and (\ref{SEB}) fulfilling the
constraints $C_{3j} =$ const at each site, and depending on a unique
frequency.
Despite the strong simplification thus introduced the complexity of
the problem is still dramatic as shown by the dynamical constraints
(\ref{PD1}) and (\ref{PD2}). It is worth recalling that their
implementation corresponds to find first the eigenvalues of
(\ref{SEA}) and (\ref{SEB}) in which $C_{3j}$ are regarded as constant,
assigned parameters, and singling out then the subset of eigenvectors
such that $|C_j|$ are compatible with the assigned $C_{3j}$. 
The staggered solutions [see Eqs. (\ref{SEO2}) and (\ref{SEO3})]
represent the case where the avalanche of initial conditions is
reduced to a set of four data namely the values of $|C_{j}|$ for the
sublattices of both even and odd sites.

Based on the polygonal symmetry of the spin equations of motion
their number has been reduced by introducing the collective variables
(\ref{collec}) in Sec. III C. The first nontrivial case (but also
the only one directly tractable in an analytic way) has been shown
to correspond to the two-sublattice solutions ($C_{j}=C_{\ell}$,
$C_{3j}=C_{3 \ell}$, $\ell= j+2$, $\forall j$).
The analysis of the fixed points of Eqs. (\ref{TSA1}) and (\ref{TSB2})
allows one to reconstruct the set of configurations in which those
corresponding to the minimum energy are implicitly contained
as a consequence of the absence of dynamics. In Sect. V we specialized
to the latter in order to obtain a zero-temperature phase space.
Noticeably, we have seen that already such a simple two-sublattice
solution contains all the qualitative features of similar diagrams
obtained in many other theories. Hence, we argue that the general
solution of fixed-point equations, if avaliable on finite lattices
by means of numerical analysis, should exhibit a richer structure
than the one obtained within standard mean-field schemes even for
what concerns the zero-temperature phase-space.

Further developments of the present work can be envisaged along the
following lines.
As to the transformation (\ref{JW}) it is important to notice how
its use has been possible because of the 1D character of the system. 
In higher dimensions, in fact, this transformation depends explicitly
on the 1D path employed to cover and thus enumerate exhaustively
the lattice sites. Such a dependence introduces in the hopping term
of the Hamiltonian a site-dependent exponential phase factor,
which does not prevent the implementation of the approach developed
here. Hence, in spite of the increased complexity thus introduced,
a natural extension of the present work is in the analysis of the
2D case dynamics.

As a matter of fact, due to the large number of degrees of
freedom involved, the 1D case itself is already not fully
tractable via numerical investigations. In this respect,
focusing on zero-dimensional systems is almost expected in
order to have a dependable numerical description. On the other hand,
it is well known that the physics of such mesoscopic systems
(e.g. quantum dots and Josephson junctions) is properly depicted
in many cases by Hubbard-like Hamiltonians \cite{QUDO}, \cite{FAZ}.
Further investigation of such systems within the scheme
proposed here seems promising. 

A final point still deserves to be deepened, which is the
requantization of the spin variables. Despite the obvious
difficulty of such a task in general \cite{BAS}, the dynamical
situations here investigated, involving the macroscopic excitation
of few system modes, seems quite feasible to this end. 
\vskip 1.0 truecm
\centerline{\bf ACKNOWLEDGMENTS}
\vskip 0.5 truecm
Part of this work was performed while the authors were visiting
the Department of Mathematics of the Open University (U.K.). The
authors would like to thank Professor A. Solomon for stimulating
discussions. One of them (V.P.) is grateful to the Open University
for its hospitality and for supporting his visit. The INFM is also
acknowledged for financial support.
\vskip 1.0 truecm
\begin{figure}
\label{fig1}
\caption{
The ground-state phase diagram of the one--band model for the
two--site solution: $d = \nu$ is the electron doping ($d =0$
half--filling) on the two--site cluster, and $1/g = U/4T$.
Its structure is in qualitative agreement with the diagram
of Ref. [10], p. 256. }
\label{fig1}
\end{figure}


\centerline{\bf APPENDIX }
\vskip 0.5 truecm
\noindent
After setting $A_j =R_{j}\exp{i \alpha_j}$,
$B_j =S_{j}\exp{i \beta_j}$,
where $R_{j} \equiv (1/4 - A^2_{3j})^{1/2}$,
$S_{j} \equiv (1/4 - B^2_{3j})^{1/2}$ with the
Poisson brackets
$\{\alpha_{\ell}, A_{3j} \} = \delta_{\ell, j}/i \hbar $,
$\{\beta_{\ell}, B_{3j} \} = \delta_{\ell, j}/i \hbar $,
it is found
\be
{\dot \alpha_j} = \delta_A - UB_{3j}
+2T A_{3j} \!
\sum_{i \in (j)} \frac{R_{i}}{R_{j}}
\cos(\alpha_{i}-\alpha_j ) ,
\label{EA1}
\ee
\be
{\dot A}_{3j} = 2T R_{j} \,
\sum_{i \in (j)} R_{i} \sin(\alpha_j -\alpha_{i}) ,
\label{EA2}
\ee
\be
{\dot \beta_j} = \delta_B - UA_{3j}
+2T B_{3j} \!
\sum_{i \in (j)} \frac{S_{i}}{S_{j}}
\cos(\beta_{i}-\beta_j ) ,
\label{EB1}
\ee
\be
{\dot B}_{3j} =2T S_{j} \sum_{i \in (j)}
S_{i} \sin(\beta_j-\beta_{i}) ,
\label{EB2}
\ee
where $(j)$ indicates the set of the nearest-neighbor sites.
%

\end{document}